# Breaking an image encryption scheme based on Arnold map and Lucas series

Imad El Hanouti[1] · Hakim El Fadili[1] · khalid Zenkouar[2]

**Abstract**

Fairly recently, a novel image encryption based on Arnold scrambling and Lucas series has been proposed in the literature [27]. The scheme design is based on permutation-substitution operations, where Arnold map is used to permute pixels for some $T$ rounds, and Lucas sequence is used to mask the image and substitute pixel's values. The authors of the cryptosystem have claimed, after several statistical analyses, that their system is "with high efficiency" and resists chosen and known plaintext attacks. Negatively, in this paper we showed the opposite. The key space of the scheme under study could be reduced considerably after our equivalent keys analysis, and thus the system is breakable under reasonable brute force attack. After all, the design of the scheme has several weaknesses that make it weak against chosen and known plaintext attacks. Consequently, we do not recommend the use of this system for any cryptographic concern or security purpose.

## 1. Introduction

With the rapid growth and development of modern telecommunication networks that pushes the limits of communication and information transmission, so many types of data have become more susceptible to be transmitted over networks without further processing or changing in their structure [1]. Multimedia data is one type of data that is widely exchanged over public networks and seldom used by public community [2-3]. A simple scenario of an eavesdropper spying on the network traffic would perceive the huge amount of proper information that could be collected about people and organizations. Thus, the security of data generally and multimedia data especially becomes a serious concern of many people and organizations.

[1] I. El Hanouti (✉) · H. El Fadili
Computer Science and Interdisciplinary Physics Laboratory (LIPI), National School of Applied Sciences, SMBA University, Fez, Morocco
Email : imad.elhanouti@usmba.ac.ma
[2] K. Zenkouar
Laboratory of Intelligent Systems and Application (LSIA), Faculty of Sciences and Technology, SMBA University, Fez, Morocco

Cryptographic protocols are one basic layer for securing data transmissions over public networks. Cryptography is, with no doubt, still the best approach to secure data and protect confidentiality of people. Modern cryptography is based on the idea of confusion-diffusion depicted by Claud Shanon in 1945 [4]. Confusion refers to making the relationship between the key and the ciphertext as complex as possible while diffusion refers to the property that the redundancy in the statistics of the plaintext is "dissipated" in the statistics of the ciphertext.

Multimedia data and especially image data by its proper intrinsic characteristics such as bulk data, strong correlation between neighboring elements and the need for fast processing algorithms [5], have motivated many researchers to adjust some existing cryptosystems to be more adequate for image cryptography [6-10]. Other authors decided to design new algorithms tailored specifically for images [11-17].

Since Arnold encrypted cat image using a chaotic map in 1967 [18], the field of chaos-cryptography has become a hot research area. Since then, so many chaotic map based encryption schemes has been proposed in the literature [14-17]. Motivated by the strong similarities between chaos theory and cryptography requirements [19], authors keep designing day after day new proposals for image and multimedia data cryptography based on chaos and other number theory aspects.

However, the majority of those new proposals have been cryptanalyzed in subsequent publications [20-23], and a crypto-game has risen up to the field of multimedia cryptology. The main reason behind the inadequacy of the proposed systems is that the majority of them are not well studied against cryptanalysis attacks (e.g. chosen/known plaintext attacks …), and so many authors rely only on some metrics and statistical analysis such that randomness tests , histogram and entropy analysis, correlation analysis, NPCR and UACI tests [24]. Consequently, many guidelines and road maps have been proposed [25-26] to ensure some sound level of security.

In [27], a new cryptosystem based on both Arnold cat map and Lucas series was proposed. The system's design is based on scrambling rounds handled by the Arnold cat map and masking process using the Lucas sequence. After several statistical tests done by the authors of [27], they claimed that their system is efficiently secure and resists the conventional known attacks like known and chosen plaintext attacks.

Negatively, our analysis has shown that the proposed system has almost no security level and should not be used for any cryptographic concern. The system is breakable under possible brute force attack after the reduction of the key space using equivalent keys analysis. The key space would never overpasses $2^{19}$ for high dimension ($1024 \times 1024$) image plaintexts. Weak keys do exist, and generally every plaintext is recoverable after some well-defined number of successive encryptions. Besides, the system is weak against known and chosen plaintext attacks. Although not discussed in this paper, the system is seemed to be weak against other attacks such as chosen ciphertext and differential attacks.

The organization of this paper is as follow: section 2 describes the system under study. Section 3 is conserved for key space analysis, brute force feasibility and weak keys analysis. Section 4 discuses a timing attack on the system. Section 5 describes two attacks: chosen plaintext attack and known plaintext attack. Finally section 6 concludes the paper.

## 2. Description of the system under study

The system under study labelled as IEAL (Image Encryption based on Arnold scrambling and Lucas series) is designed for grayscale and color square[3] images. However, in this paper we suppose that all images are 8-bit grayscale square image with size $N \times N$ i.e. image= $\{I(i,j)\}_{i=0,j=0}^{N-1,N-1}, I(i,j) \in \{0,1,\ldots,255\}$. The cryptanalytic results are extendable to any image color depth.

IEAL key exchanging is carried out using public key encryption algorithm. The key[4] consists of two integer numbers: the first denoted as $T$ is for scrambling round iterations, and the last one is for the first position of Lucas sequence denoted as $S$. After Diffie-Hellman based key exchanging, IEAL encryption process could be described briefly in two main steps:

1) *Image scrambling*:
   Taken an input image with size $N \times N$ (denoted in this paper as the plaintext/plain image $I$), $I$ is scrambled $T$ times according to Arnold cat map [18] as follow:
   $$\begin{pmatrix} i^* \\ j^* \end{pmatrix} = \begin{pmatrix} i+j \ (mod \ N) \\ i+2j \ (mod \ N) \end{pmatrix} \quad (1)$$
   Where $i^*$ and $j^*$ are the new pixel coordinates after each iteration, $i$ and $j$ are the old pixel coordinates, $mod$ is denoting modular arithmetic. The scrambled image could be formulated in function of the plain image as:
   $$I^*(i^*, j^*) = I(i,j) \quad (2)$$
   Where $i^* = A_T(i)$ and $j^* = A_T(j)$, $A_T$ denotes the application of Eq.(1) $T$ times.

2) *Image masking*:
   The image $I^*$ obtained after scrambling stage is masked pixel-by-pixel in raster order with a sequence[5] $s = \{s(k)\}_{k=0}^{N.N-1}$ obtained from Lucas sequence starting its generation from the position $s$ as follow:
   $$s(k) = L_{S+k} \ (mod \ 256), \quad k = 0 \sim N \times N - 1 \quad (3)$$
   Where $L_n$ is Lucas sequence defined as :
   $$L_n = \begin{cases} 2, & n = 0 \\ 1, & n = 1 \\ L_{n-1} + L_{n-2}, & n > 1 \end{cases} \quad (4)$$
   Thus, we formulate the encrypted image as:
   $$I'(i,j) = I^*(i,j) \oplus s(i \times N + j), \quad i,j = 1 \sim N \quad (5)$$

The all encryption process could be formulated mathematically as follow:

$$I'(i^*, j^*) = I(i,j) \oplus s(i^* \times N + j^*), \quad i,j = 1 \sim N \quad (6)$$

Where $i^* = A_T(i)$ and $j^* = A_T(j)$, $A_T$ denotes the application of Eq.(1) $T$ times.

## 3. Key space analysis

---

[3] Although not explicitly mentioned, the authors of IEAL use N to indicate the image size and N.N to indicate number of pixels. For convenience with Arnold map, the image may have to be square.
[4] Some notations from the original paper are modified without affecting its main meaning.
[5] IEAL's authors did not mention how to process Lucas series for XOR masking; however we reduce the sequence modulo 256 to maintain pixels depth.

As mentioned in the previous section, the key is composed of two integer numbers $(T, S)$ for iterations and starting position of Lucas sequence generation respectively. No key space analysis has been done in the IEAL's main article. Our analysis shows that the key space of the IEAL system is very weak.

### 3.1 Equivalent keys

From first view, the numbers $T$ and $S$ of iterations and starting position respectively could be any integer numbers. However, the system of Eq.(1) has been shown to be periodic [28], and thus, $T$ itself is periodic. That is, if $m$ was the period of the system of Eq.(1), then for any secret key $T$, $T + m$ is an equivalent key. Thus, $T$ could take only $m$ different values. So that the secret key $T$ is within a pre-defined set $\{0,1,2,\ldots,m-1\}$ where $m$ is the *period* of the map in Eq.(1). Many studies [28,30] have shown that $m$ has an upper bound in function of the size $N$ of the image as follow:

$$m \begin{cases} = 3N, & \text{for } N = 2 \times 5^r \\ = 2N, & \text{for } N = 5^r \text{ or } N = 6 \times 5^r \quad , r \in \mathbb{N} \\ \leq \frac{12}{7} N, & \text{for other } N \end{cases} \qquad (7)$$

***Proposition 1***: for any image of size $N \times N$, and for any key (number of iterations) $T$, all the keys that belongs to the set $K^T = \{T + n \times m \mid n \in \mathbb{N}\}$, are equivalent keys of $T$. Where $m$ is defined in Eq.(7).

In the other hand, calculations on the sequence $s$ of Eq.(3) has shown that $s$ is periodic by its turn. The *period* of the sequence $s$ after some number of calculations is found to be $p = 384$.

***Proposition 2:*** We conjecture that the Lucas sequence modulo $2^8$ is *periodic-384*. Thus, for any key $S$ (start position of the sequence), all the keys of the set $K^S = \{S + 384.n \mid n \in \mathbb{N}\}$, are equivalent keys.

### 3.2 Key space and brute force attack

From propositions 1 and 2, the actual space of keys $Ks$ for a brute force attack could be minimized considerably in function of the dimension size $N$ of image as follow:

$$Ks \begin{cases} = 1152.N, & \text{for } N = 2 \times 5^r \\ = 768.N, & \text{for } N = 5^r \text{ or } N = 6 \times 5^r \quad , r \in \mathbb{N} \\ \leq \frac{4608}{7} N, & \text{for other } N \end{cases} \qquad (8)$$

For $1024 \times 1024$ high dimension images, the key space for brute force attack would never overpasses $2^{19.3625}$. However, as depicted in Table 1 this theoretical upper bound value is greater than the real value. This is considered as a very weak key space for brute force attack as the minimum key space recommended being not less than $2^{128}$ [26]. Even more, some calculations depicted on Table 1 of the key space length according to some image sizes shown that the actual key space size would be in some cases very small than the upper bound formulated in Eq.(8).

**Table 1** Key space size according to some image sizes.

| Image size $N \times N$: | The key space for brute force attack: |
|---|---|
| $N = 124$ | $Ks = 15 \times 384 = 2^{12.49}$ |
| $N = 128$ | $Ks = 96 \times 384 = 2^{15.17}$ |
| $N = 144$ | $Ks = 12 \times 384 = 2^{12.17}$ |
| $N = 256$ | $Ks = 192 \times 384 = 2^{16.17}$ |
| $N = 276$ | $Ks = 24 \times 384 = 2^{13.17}$ |
| $N = 300$ | $Ks = 300 \times 384 = 2^{16.81}$ |
| $N = 341$ | $Ks = 15 \times 384 = 2^{12.49}$ |
| $N = 377$ | $Ks = 14 \times 384 = 2^{12.39}$ |
| $N = 512$ | $Ks = 384 \times 384 = 2^{17.17}$ |
| $N = 1024$ | $Ks = 768 \times 384 = 2^{18.17}$ |

After some computer calculations, and without affecting any cryptanalysis results in this paper, we conjecture that for any image size of $2^n \times 2^n$ with $n > 2$ the key space size would be $= 384 \times 3 \times 2^{n-2} = 3^2 \times 2^{n+5} \simeq 2^{n+8.1699}$ .

After our key space analysis, and considering the tiny structure of the cryptosystem (upper bound of complexity $O(n.T)$ with $n$ is the size of the image $n = N \times N$), we could claim that the system IEAL is extremely weak under brute force attack. In Fig. 1 we show the result of a brute force attack on an unknown $144 \times 144$ cipher image encrypted using randomly generated pair of keys $(T, S) = (13, 390)$. Note that all keys used for encryption and illustrations in this paper are generated using *rand* function from *octave*[6], and that all images used in this paper for illustration purpose are from SIPI[7] database. The brute force attack ends after 29.454 seconds recovering the original plaintext with an equivalent key $(Se, Te) = (6, 1)$. That is, $390 \ (mod \ 384) = 6 \ and \ 13 \ (mod \ 12) = 1$.

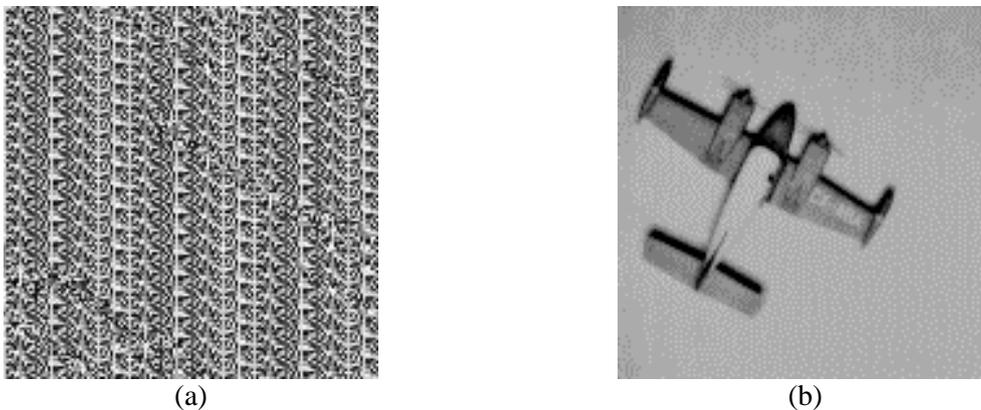

(a)            (b)

**Fig. 1** The proposed brute force attack. (a) cipher image of size $144 \times 144$, (b) recovered plain image after 29.4 seconds

### 3.3 Weak keys

In cryptology, weak keys are keys that lead to some undesired results (e.g. no or *bad* encryption, self-decryption…). The issue arises when the probability of picking up a weak key randomly is not negligible. However, for any cryptographic system, it is better worth to exclude all weak keys from the key space.

In this approach, if re-encrypting the same image with the same key leads to decrypting the image, thus, the key pair is considered as a weak key. For any key $S$ ( start position), if the key $T$ is chosen to be in the form $n.m, n \in \mathbb{N}$, where $m$ is the period as defined above in Eq.(7), then the pair $(T,S)$ is a weak key, because after two encryptions with the same key we will reproduce the original image. According to Eq.(6) the first encrypted image is :

$$I^1(i^1, j^1) = I(i,j) \oplus s(i^1 \times N + j^1) \tag{9}$$

---

[6] https://gnu.org/software/octave/
[7] collection of digitized images at http://sipi.usc.edu/database/

Since there is no scrambling because $T = n.m$ is equivalent to zero according to proposition 1, and thus no scrambling is done on the original image, only masking operation remains:

$$I^1(i,j) = I(i,j) \oplus s(i \times N + j) \tag{10}$$

For the second encryption, another cipher image is obtained as:

$$I^2(i,j) = I^1(i,j) \oplus s(i \times N + j) \tag{11}$$

By substituting $I^1$ in Eq.(11) we obtain:

$$I^2(i,j) = I(i,j) \oplus s(i \times N + j) \oplus s(i \times N + j) = I(i,j) \tag{12}$$

Thus, the original image is obtained by re-encrypting the encrypted one. Consequently, the pair of this particular key is considered as weak key. The probability of picking up a weak key is $Pr = 1/m$, where $m$ is defined in Eq.(7) (e.g. for $144 \times 144$ size image the probability of picking up a weak key is of 8.33% which is not negligible at all).

Fig. 2 illustrates this type of weakness, after re-encrypting the $276 \times 276$ size cipher image encrypted with the randomly generated key $= 202$, and a chosen $T = 24 \times 2 = 48$ we recovered the original plain image by simple application of a second encryption to the cipher image.

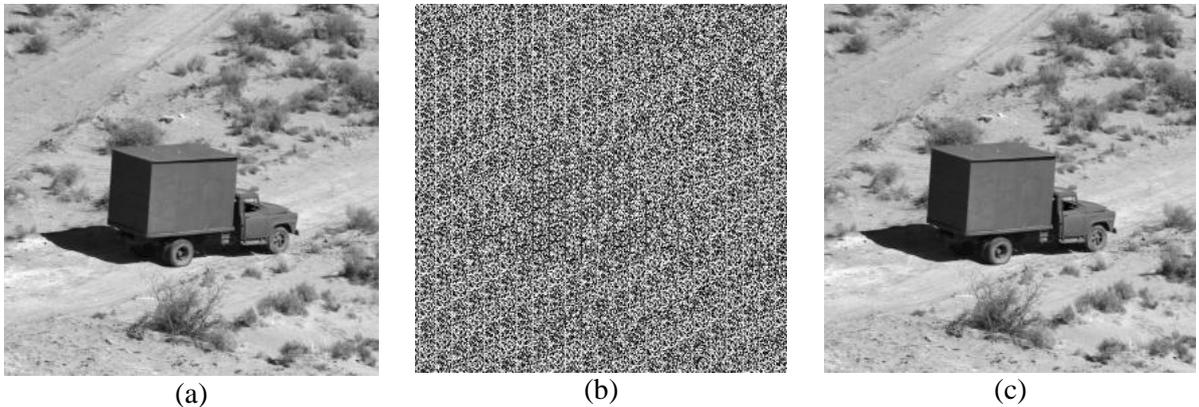

(a)          (b)          (c)

**Fig. 2** Illustration of a weak key encryption (a) original image, (b) cipher image after the first encryption of the original image, (c) cipher image after the second encryption of the original image which is the original one.

### 3.3.1 Generalized cycle attack

Let us consider that an attacker could perform any number of re-encryptions for a particular unknown plaintext using the same unknown pair of key $(T,S)$. This typical attack could be seen as a type of chosen plaintext attack.

**Proposition 3:** Only $n = \frac{2m}{\gcd(T,m)}$ re-encryption is needed to obtain the original plain image by the generalized re-encryption attack, where $m$ is the period as defined in Eq.(7) and $\gcd(x, y)$ is the greatest common divisor of integer numbers $x$ and $y$.

**Proof:** trivially proved from the fact that given a number of iterations $T$, and the period of iteration is $m$, thus, one needs $r = \frac{lcm(T,m)}{T}$ another iteration to obtain the original image, where $lcm(x, y)$ is the least common multiple of two integers $x$ and $y$.

In order to dissipate masking effect, we need another $r$ application of the encryption process.

Thus, we need $n = \frac{lcm(T,m)}{T} \times 2$ successive encryptions to recover the original plain image.

Since for any two positive integers $T$ and $m$ we have: $lcm(T, m) \times \gcd(T, m) = mT$, we finally get:

$$n = \frac{lcm(T,m)}{T} \times 2 = \frac{mT}{T.\gcd(T,m)} \times 2 = \frac{2m}{\gcd(T,m)} \quad (13)$$

□

This attack allows us to recover not only the plain image but also to recover keys after enhancing the brute force attack to a dictionary attack by limiting the values that the key $T$ could take for scrambling process according to the following equality:

$$\gcd(T, m) = \frac{2m}{n} \quad (14)$$

As for illustration purpose, we consider an $144 \times 144$ cipher image shown in Fig. 3. By application of successive re-encryptions we end up by recovering the plain image after 24 encryptions as shown in Fig.3. From Table 1 we have $m = 12$. From Eq.(14) we have: $\gcd(T, 12) = \frac{2 \times 12}{24} = 1$. Since from proposition 1 $T \in \{0,1,..,11\}$ we conclude that $T \in \{1,5,7,11\}$, that is, the secret key used for encryption in Fig.3 has been chosen randomly to be $(T, S) = (11,68)$.

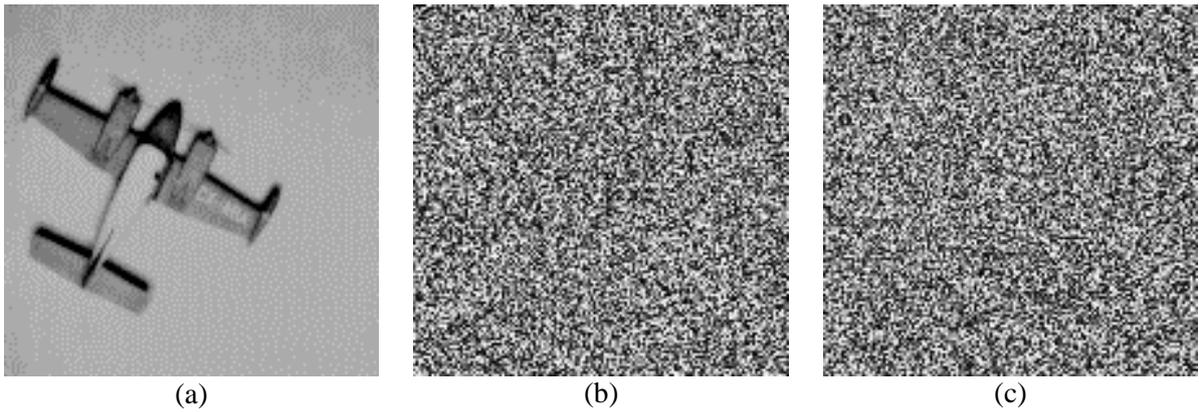

(a)          (b)          (c)

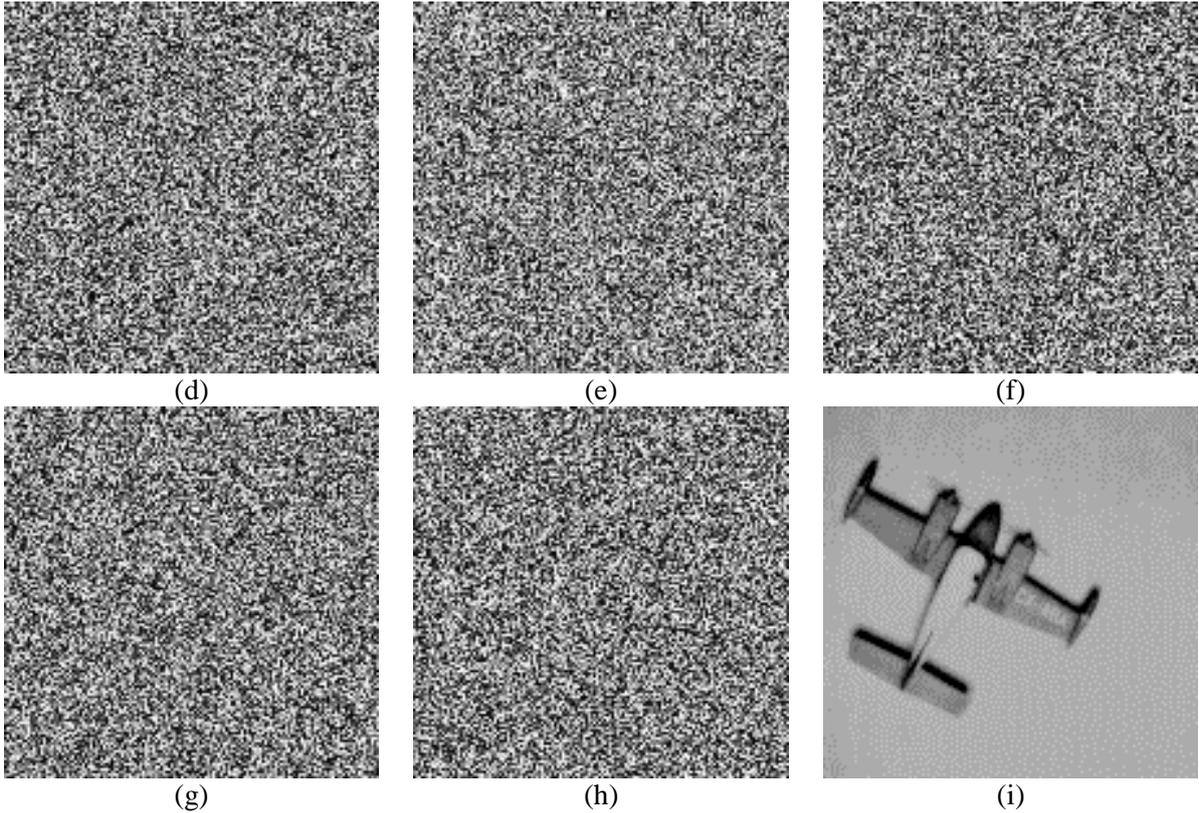

**Fig. 3** generalized cycle attack (a) the original image (b) cipher image after 3 encryptions (c) after 6 encryptions (d) after 9 encryptions (e) after 12 encryptions (f) after 15 encryptions (g) after 18 encryptions (h) after 21 encryptions (i) after 24 encryptions which is the same plain image.

### 4. Timing attack

In this section, we discuss the applicability of a side-channel attack based on time consumption measure. Thus, we consider that an attacker has the ability to access or obtain some statistical analysis of timing measurements of the cryptosystem for encrypting some unknown plaintexts with unknown keys.

For a given data inputs: plain image with size $n = N \times N$ and a pair of secret key $(T, S)$, assuming no optimizations are done on the encryption algorithm, the complexity of the algorithm could be of:

$$O(n.T + S + n) \simeq O(n(T + 1)) \qquad (15)$$

Since the time execution increases with number of iterations, and that only one masking operation is done whatever the key is. The number of iterations could be considered as the only parameter that affects time execution metrics. Actually, timing attack is more dependent on the processor design and the code optimization, but since the number of iterations constitutes a half-key, this attack could be successfully exploited to reveal half of the key or to enhance considerably a brute force attack by setting up a dictionary attack on $T$.

### 5. Chosen plaintext and known plaintext attack

In [27] the authors of IEAL system have declared that their proposed system resists chosen/known attacks, negatively, this claim is not true. In the next two subsections we discuss the weakness of IEAL against chosen and known plaintext attacks. Moreover we could claim that the IEAL cryptosystem is weak also under several other types of attacks like chosen ciphertext and differential attack.

### 5.1 chosen plaintext attack

In a chosen plaintext attack scenario, the attacker has temporary access to the encryption machinery, and he could choose some intentionally designed plaintexts to be encrypted using unknown key. In our case of study, we will demonstrate that the maximum required plaintexts to recover an equivalent key of the encryption are $\lceil 2.log_{256}(N) \rceil + 1$ chosen plaintexts where $\lceil . \rceil$ denotes the *ceil* function, that is the smallest integer not smaller than (.).

- *Recovering the masking key:*
  To recover the masking Lucas sequence, we have first to get ride off scrambling process by introducing an all-zero pixel plain image $I_0 = [I_0(i,j)]$ where $I_0(i,j) = 0$ for $i,j = 0 \sim N-1$ and $N \times N \geq 384$. After encrypting $I_0$ we will get a cipher image $I'_0$ with blocks of the equivalent key since from Eq.(6) we have :
  $$I'_0(i,j) = I_0(i,j) \oplus s(i \times N + j) = s(i \times N + j), \; i,j = 1 \sim N \quad (16)$$
- *Recovering the scrambling map:*
  After recovering the masking key, the scheme could be seen as a permutation-only encryption scheme, thus, as demonstrated in [29], only a number of $\lceil log_{256}(N \times N) \rceil$ of chosen plaintexts is required to recover the scrambling equivalent key.

Thus, the total number of required chosen plaintexts to recover the full equivalent key for encryption is $\lceil 2.log_{256}(N) \rceil + 1$. As for $1024 \times 1024$ image size, we need only 3 chosen plaintexts to recover an equivalent key.

### 5.2 Known plaintext attack

For known plaintext attack, the attacker knows some plaintexts and their corresponding ciphertexts. The attacker would try to recover the unknown key or part of it. If this is done, then the system would be considered as unsecure.

From this perspective and by exploiting the attack described in the subsection 3.3, we could recover the pair of keys $(T, S)$ by a dictionary attack made up by exploiting information from one pair of known plaintext/ciphertext. Consider that we have pair of plaintext/ciphertext denoted as $I$ and $C$.

- *Recovering the key/equivalent key of $S$:*
  First of all, let us determine if any fixed points exist in the Arnold map, by solving the following system:
  $$\begin{cases} i + j \; (mod \; N) = i \\ i + 2j \; (mod N) = j \end{cases}, i,j = 0,1,2..N-1 \quad (17)$$
  Obviously, the only fixed point is the point (0,0). Hence, the pixel with coordinates (0,0) would never changes its position whatever the key $T$ was. The first element of the sequence $s$ could be obtained as:
  $$s(0) = I(0,0) \oplus C(0,0) \quad (18)$$
  Since $s$ contains same elements but those elements are shifted dependently on the value of $S$ key, the position of $s(0)$ will narrow up the size of the dictionary of $S$ key. Table 2 shows that the size of dictionary for $S$ key could take only 4 values depending on the value of $s(0)$. The probability of using each size value is also depicted in Table 2. Measures shown in table 2 are simply calculated from the statistical analysis of the *384-length* Lucas sequence used in the IEAL cryptosystem. For each value of probable key $\hat{S}$, we decrypt masking-only operation and we compare the histogram of the encrypted image $\hat{C}$ with the histogram of the plain image $I$. If the two histograms matches, then the key (or equivalent key of $S$) is $S = \hat{S}$.

- *Recovering the key/equivalent key of $T$:*
  After the recovery of the key $S$, we could perform the cycle attack on an scrambled-only version of the plain image which is $\hat{C}$ using one iteration for each encryption $\hat{T} = 1$. Since no masking process is done, and after recovering the plain image, say after $n$ scrambling-only encryption, the equivalent key for $T$ could be recovered as: $T = m - n$. Where $m$ is defined in Eq.(7). This result could be proved easily by the fact that $m$ scrambling is equivalent to zero, thus $T + n$ is equivalent to zero. Hence, $m - n$ is the key $T$ or an equivalent key of it.

**Table 2** size of dictionary attack for $S$ key using known plaintext attack and the probability of using each dictionary size.

| Size of dictionary for $S$ key using known plaintext attack: | Probability of using this dictionary size: |
|---|---|
| 1 | $pr = {}^{64}/_{384} = 0.1666$ |
| 2 | $pr = {}^{64}/_{384} = 0.1666$ |
| 3 | $pr = {}^{192}/_{384} = 0.5$ |
| 16 | $pr = {}^{64}/_{384} = 0.1666$ |

Fig. 4 illustrates known plaintext attack using $276 \times 276$ size images. The objective is to recover the key (or equivalent key) which is supposed to be unknown. The known plaintext is shown in Fig. 4(a), its corresponding known ciphertext is shown in Fig. 4(b). The main attack's process is demonstrated in the next few steps:

- Calculation of $s(0)$ according to Eq.(18): $s(0) = I(0,0) \oplus C(0,0) = 109$.
- The dictionary size is $d = 1$, since $s(0)$ exists only once in the 384-length Lucas sequence.
- Calculation of $S$ gives, $S = 127$.
- Decryption of masking-only process of $C$ produces a scrambled-only image $\hat{C}$ shown in Fig. 4(c).
- Successive scrambling encryptions of the intermediate cipher image $\hat{C}$ leads to the recovery of the plain image $P$ after $n = 18$ encryptions. The recovered image is shown in Fig. 4(d).
- The calculation of $T$ gives $T = m - n = 6$. Note that from table 1, we have $m = 24$.
- The recovered *equivalent* key is $(T, S) = (6, 127)$, that is, the original key was $(T, S) = (6, 127)$.
- Use the recovered key to decrypt the cipher image shown in Fig. 4(e).
- Decrypted image is shown in Fig. 4(f) which is exactly the unknown plain image.

Note that this method is not the only (nor maybe the best) method for implementing a known plaintext attack against the IEAL system. The method depicted in this subsection is to prove the weakness of the IEAL system against known plaintext attack.

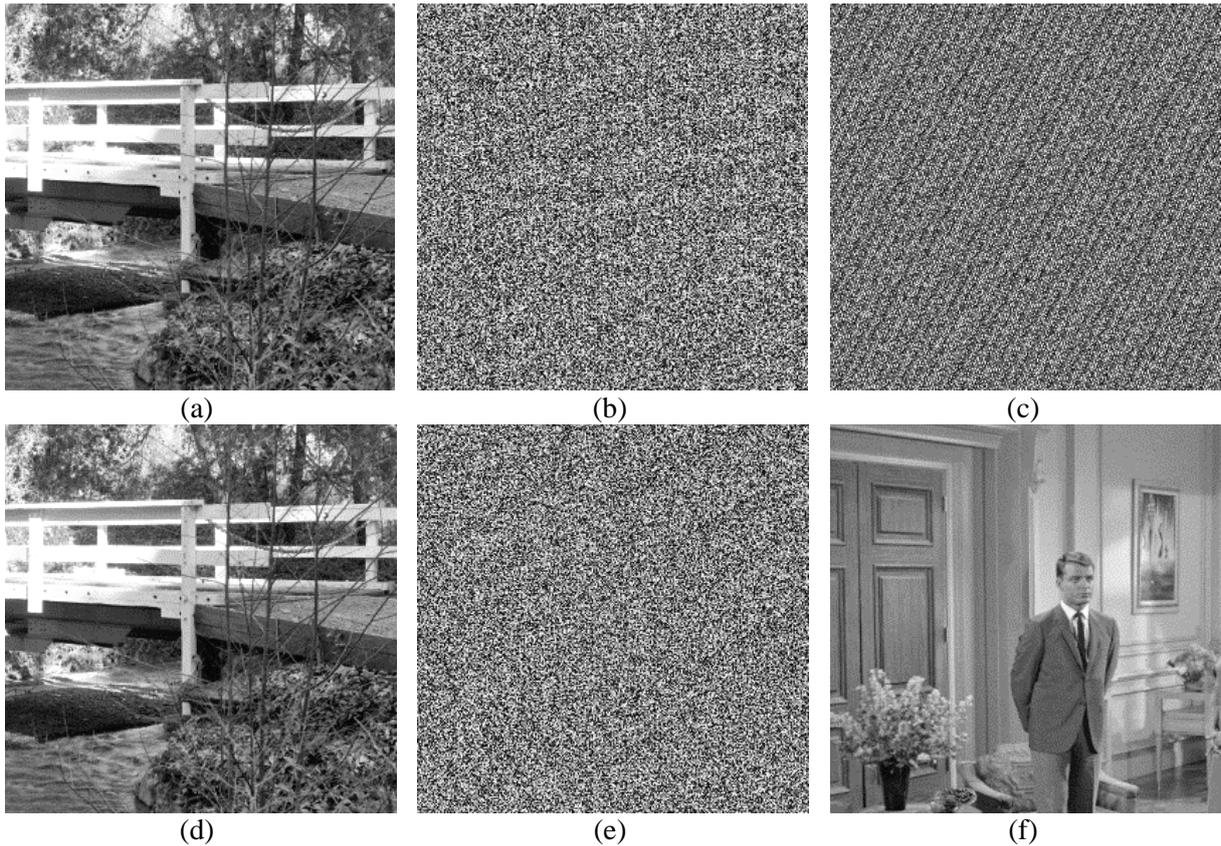

**Fig 4** the proposed known plaintext attack (a) the known plaintext (b) the known ciphertext (c) scrambled only cipher (d) the encountered plaintext (e) cipher text of unknown image (f) the recovered image

## 6. Conclusion

In this paper, we scrutinize the design of a recently proposed cryptosystem for image encryption based on Arnold scrambling and Lucas series masking. The scheme under study is proved in this paper to be inappropriate for cryptographic use. Yet, the underlined algorithm exhibits bad secrecy insurance and many imperiled typical algorithmic flaws. Although the system did pass several statistical tests, it cannot resist conventional attacks such as chosen plaintext and known plaintext attacks. Even worst, the system is breakable under feasible brute force attack after the reduction of the key space by our equivalent keys analysis. In addition to that, the existence of many weak keys and its vulnerability against cycle attack (successive encryptions) completely makes the cryptosystem faulty and with no sound secrecy. Lastly, we do not recommend the use of this cryptosystem for any information security and data protection concern.